\documentclass[final,times,5p]{elsarticle}
\usepackage{ctable}
\usepackage{graphicx}
\usepackage{xspace}
\usepackage{hhline}
\usepackage{amsmath}
\usepackage{amsfonts}
\usepackage{hyperref}

\DeclareMathOperator{\Imag}{Im}
\newcommand{\ee}{\ensuremath{e^{+}e^{-}}\xspace}
\newcommand{\mumu}{\ensuremath{\mu^{+}\mu^{-}}}

\newcommand{\JP}{\ensuremath{J/\psi}\xspace}

\renewcommand{\epsilon}{\varepsilon}

\begin{document}
\title{Measurement of $R$ between 1.84 and 3.05 GeV at the KEDR detector}
\author[binp]{V.V.~Anashin}
\author[binp,nsu]{V.M.~Aulchenko}
\author[binp,nsu]{E.M.~Baldin} 
\author[binp]{A.K.~Barladyan}
\author[binp,nsu]{A.Yu.~Barnyakov}
\author[binp,nsu]{M.Yu.~Barnyakov}
\author[binp,nsu]{S.E.~Baru}
\author[binp]{I.Yu.~Basok}
\author[binp]{A.M.~Batrakov}
\author[binp,nsu]{A.E.~Blinov}
\author[binp,nsu,nstu]{V.E.~Blinov}
\author[binp,nsu]{A.V.~Bobrov}
\author[binp,nsu]{V.S.~Bobrovnikov}
\author[binp,nsu]{A.V.~Bogomyagkov}
\author[binp,nsu]{A.E.~Bondar}
\author[binp,nsu]{A.R.~Buzykaev}
\author[binp,nsu]{S.I.~Eidelman}
\author[binp,nsu,nstu]{D.N.~Grigoriev}
\author[binp]{Yu.M.~Glukhovchenko}
\author[binp]{S.E.~Karnaev}
\author[binp]{G.V.~Karpov}
\author[binp]{S.V.~Karpov}
\author[binp]{P.V.~Kasyanenko}
\author[binp]{T.A.~Kharlamova}
\author[binp]{V.A.~Kiselev}
\author[binp]{V.V.~Kolmogorov}
\author[binp,nsu]{S.A.~Kononov}
\author[binp]{K.Yu.~Kotov}
\author[binp,nsu]{E.A.~Kravchenko}
\author[binp,nsu]{V.N.~Kudryavtsev}
\author[binp,nsu]{V.F.~Kulikov}
\author[binp,nstu]{G.Ya.~Kurkin}
\author[binp]{I.A.~Kuyanov}
\author[binp,nsu]{E.A.~Kuper}
\author[binp,nstu]{E.B.~Levichev}
\author[binp,nsu]{D.A.~Maksimov}
\author[binp]{V.M.~Malyshev}
\author[binp,nsu]{A.L.~Maslennikov}
\author[binp,nsu]{O.I.~Meshkov}
\author[binp]{S.I.~Mishnev}
\author[binp,nsu]{I.I.~Morozov}
\author[binp,nsu]{N.Yu.~Muchnoi}
\author[binp]{V.V.~Neufeld}
\author[binp]{S.A.~Nikitin}
\author[binp,nsu]{I.B.~Nikolaev}
\author[binp]{I.N.~Okunev}
\author[binp,nsu,nstu]{A.P.~Onuchin}
\author[binp]{S.B.~Oreshkin}
\author[binp,nsu]{A.A.~Osipov}
\author[binp,nstu]{I.V.~Ovtin}
\author[binp,nsu]{S.V.~Peleganchuk}
\author[binp,nstu]{S.G.~Pivovarov}
\author[binp]{P.A.~Piminov}
\author[binp]{V.V.~Petrov}
\author[binp,nsu]{V.G.~Prisekin}
\author[binp,nsu]{O.L.~Rezanova}
\author[binp,nsu]{A.A.~Ruban}
\author[binp]{V.K.~Sandyrev}
\author[binp]{G.A.~Savinov}
\author[binp,nsu]{A.G.~Shamov}
\author[binp]{D.N.~Shatilov}
\author[binp,nsu]{B.A.~Shwartz}
\author[binp]{E.A.~Simonov}
\author[binp]{S.V.~Sinyatkin}
\author[binp]{A.N.~Skrinsky}
\author[binp,nsu]{A.V.~Sokolov}
\author[binp,nsu]{A.M.~Sukharev}
\author[binp,nsu]{E.V.~Starostina}
\author[binp,nsu]{A.A.~Talyshev}
\author[binp,nsu]{V.A.~Tayursky}
\author[binp,nsu]{V.I.~Telnov}
\author[binp,nsu]{Yu.A.~Tikhonov}
\author[binp,nsu]{K.Yu.~Todyshev \corref{cor}}
\cortext[cor]{Corresponding author, e-mail: todyshev@inp.nsk.su}
\author[binp]{G.M.~Tumaikin}
\author[binp]{Yu.V.~Usov}
\author[binp]{A.I.~Vorobiov}
\author[binp,nsu]{V.N.~Zhilich}
\author[binp,nsu]{V.V.~Zhulanov}
\author[binp,nsu]{A.N.~Zhuravlev}
 \address[binp]{Budker Institute of Nuclear Physics, 11, akademika
 Lavrentieva prospect,  Novosibirsk, 630090, Russia}
 \address[nsu]{Novosibirsk State University, 2, Pirogova street,  Novosibirsk, 630090, Russia}
 \address[nstu]{Novosibirsk State Technical University, 20, Karl Marx
  prospect,  Novosibirsk, 630092, Russia}

\begin{abstract} 
Using the KEDR detector at the VEPP-4M \ee collider, we have
determined the values of $R$ 
at thirteen points of the center-of-mass energy between 1.84 and 3.05
GeV. The achieved accuracy is about or better than  $3.9\%$ at
most of the energy points  with a systematic uncertainty less than $2.4\%$.
\end{abstract}
\maketitle
\section{Introduction}\label{sec:intro}
Measurement of the $R$ value has long history and became a classical
experiment on high energy  physics. The quantity $R$ is defined as
\begin{equation}
R=\frac{\sigma(\ee\!\to\!\text{hadrons})}{\sigma(\ee\!\to\!\mumu)},
\end{equation}
where $\sigma(\ee\!\to\!\text{hadrons})$ is the radiatively-corrected 
total \\  hadronic cross section in electron-positron annihilation and \\
$\sigma(\ee\!\to\!\text{\mumu})$ is the lowest-order QED cross section  of the muon pair production. 

The experiments devoted to the $R$ measurement in the energy range from 1.8
GeV up to the vicinity of $J/\psi$ are described in 
Refs.\cite{Mark1:R1977,GG2:R1979,ADONEMUPI:R1973,ADONE:R1981,Bai:1999pk,BES:R2002,BES:R2009}. 
The accuracy of BES-II results~\cite{BES:R2009} 
at 2.6 and 3.07 GeV reaches 3.8\% and 3.3\%, respectively, while the precision in other
experiments does not exceed $5\%$.

Precise measurements of the $R(s)$ dependence  play an important role in the
determination of the running strong  coupling constant
$\alpha_s(s)$ and heavy  quark masses~\cite{quark},
the anomalous magnetic moment of the muon $(g-2)_\mu$ and the value of the
electromagnetic fine structure constant at the $Z^0$ peak
$\alpha(M_Z^2)$~\cite{dhmz,hlmnt}. 
A significant contribution to uncertainties of the quantities 
listed above comes from the energy region below charm threshold, 
which motivated us to perform new $R(s)$ measurements.  

KEDR has recently published the $R$ values at seven points of the
center-of-mass energy between 3.12 and 3.72~GeV \cite{KEDR:R2016}. 
In this paper we present $R(s)$ measurements in the energy range 
from 1.84 GeV up to 3.05 GeV. The experiment with an integrated 
luminosity of about 0.66~pb$^{-1}$ was carried out in 2010.
Our result considerably improves the existing $R(s)$ measurements in this
energy range and would be useful for matching CMD-3 and SND data which will be
obtained by summing cross sections of the exclusive modes.

\section{VEPP-4M collider and KEDR detector}\label{sec:VEPP}
The  \ee collider VEPP-4M ~\cite{Anashin:1998sj}  can operate in the
2$\times$2 bunches mode in the wide range of the beam energy. 
The peak luminosity of VEPP-4M is 
about~$10^{30}\,\text{cm}^{-2}\text{s}^{-1}$  in the vicinity of \JP 
and drops to $10^{29}\,\text{cm}^{-2}\text{s}^{-1}$ at the beam energy of 1~GeV.

The  VEPP-4M is equipped with two systems of beam energy calibration. 
The resonant depolarization method \cite{Bukin:1975db,Skrinsky:1989ie}
is used for precise mass measurements \cite{UFN:KEDR2014,MASS::KEDR2015}.
In experiments requiring long-term data collection the 
energy monitoring is performed with the infrared light Compton
backscattering (CBS) \cite{CBS}.

A detailed description of the KEDR detector can be found in
Ref.~\cite{KEDR:Det}. Charged particles are reconstructed by 
the drift chamber (DC) and vertex detector (VD) which compose the
tracking system of the detector. 
Electrons are identified by the ratio of the energy deposited 
in the CsI and LKr calorimeters to the track momentum.
The particle identification system is based on the aerogel Cherenkov counters. 
The primary trigger (PT) operates using signals from
the time-of-flight (TOF) counters and fast signals from the CsI and
LKr calorimeters, the secondary trigger (ST) uses optimally shaped
calorimeter signals and the information from the VD, DC and TOF
systems~\cite{TALYSHEV}.
Muons are identified in the muon system inside the magnet yoke.
The superconducting solenoid provides a longitudinal magnetic field of 0.6~T.
The detector is equipped with a tagging system of scattered electrons  
for two-photon studies. The on-line luminosity measurement is provided
by two independent single bremsstrahlung monitors.

\section{Experiment}\label{sec:exp}
The purpose of the experiment was the determination of the total hadron
cross section at thirteen equidistant points between 1.84 and 3.05 GeV.
During data taking there were some problems with the laser for
CBS energy measurements.
At most points the energy was determined using the correction of
the calculated accelerator energy. These corrections were found
in the experiment on the narrow resonance search~\cite{KEDR:NarrRes2011}.
The accuracy of beam energy determination was about 1~MeV that was checked
using a few CBS calibrations performed during the $R$ scan.

The actual energy and integrated luminosity at all 
points are presented in Table~\ref{tab:epoints}.
The systematic uncertainty in the measured integrated luminosity
is considered in Sec.~\ref{sec:lumerr}.

\renewcommand{\arraystretch}{1.7}
\setlength{\tabcolsep}{3pt}
\begin{center}
\begin{table}[h!]
\begin{center}
\caption{{\label{tab:epoints} Center-of-mass energy $\sqrt{s}$ and
    integrated luminosity $\int\!\!\mathcal{L}dt$} in the R scan points.} 
\begin{tabular}{lcc}

Point & $\sqrt{s}$, MeV &  $\int\!\!\mathcal{L}dt$, nb$^{-1}$ \\\hline
1& $1841.0  $   &  ~~$ 10.32\pm 0.19 \pm 0.12$ \\ \hline
2& $1937.0 $   &  ~~$ 29.13\pm 0.34 \pm 0.35$ \\ \hline
3& $2037.3 $   &  ~~$ 43.16\pm 0.44 \pm 0.52$ \\ \hline
4& $2135.7 $   &  ~~$ 43.29\pm 0.46 \pm 0.52$ \\ \hline
5& $2239.2 $   &  ~~$ 46.40\pm 0.49 \pm 0.56$ \\ \hline    
6& $2339.5 $   &  ~~$ 54.55\pm 0.56 \pm 0.65$ \\ \hline
7& $2444.1  $  &  ~~$ 52.80\pm 0.57 \pm 0.63$ \\ \hline   
8& $2542.6  $  &  ~~$ 52.13\pm 0.59 \pm 0.63$ \\ \hline
9& $2644.8  $  &  ~~$ 55.43\pm 0.64 \pm 0.67$ \\ \hline
10& $2744.6 $  &  ~~$ 66.80\pm 0.72 \pm 0.80$ \\ \hline
11& $2849.7 $  &  ~~$ 69.14\pm 0.77 \pm 0.83$ \\ \hline
12& $2948.9 $  &  ~~$ 75.87\pm 0.83 \pm 0.91$ \\ \hline    
13& $3048.1 $  &  ~~$ 60.08\pm 0.76 \pm 0.72$ \\ 
\end{tabular} 
\end{center}
\end{table}
\end{center}
\renewcommand{\arraystretch}{1.}
\section{Data analysis}\label{sec:data}
\subsection{Analysis procedure}\label{subsec:proc}
The observed hadronic annihilation cross section 
was determined from
\begin{equation}
\sigma_{\text{obs}}(s)=\frac{N_{\text{h}} - N_{\text{res.bg.}}}{\int\!\!\mathcal{L}dt},
\label{eq:sigmaobs}
\end{equation}
where $N_{\text{h}}$ is the number of events that meet hadronic
selection criteria,
$N_{\text{res.bg.}}$ is the residual machine background evaluated as
 discussed in Sec.~\ref{sec:background},
and $\int\!\!\mathcal{L}dt $ is the integrated luminosity.

For the given observed cross section, the $R$ value was calculated
as follows:
\begin{equation}
\label{eq:R}
R = \frac{\sigma_{\text{obs}}(s)-
\sum\epsilon_{\text{bg}}(s)\,\sigma_{\text{bg}}(s)} {\epsilon(s)\, (1+\delta(s)) \,\sigma_{\mu\mu}(s)},
\end{equation}
where $\sigma_{\mu\mu}(s)=4\pi\alpha^2/3s$ is the Born cross section
for $\ee\!\to\!\mumu$, and
$\epsilon(s)$ is the detection efficiency for the single photon annihilation
to hadrons. The second term in the numerator corresponds to the physical
background from  $e^+e^-$, $\mu^+\mu^-$ production and two-photon processes.
The radiation correction factor $1+\delta(s)$ can be written as
\begin{equation}
\label{eq:RadDelta}
1+\delta(s)=\int\!\frac{dx}{1\!-\!x}\, 
\frac{\mathcal{F}(s,x)}{\big|1-\Pi((1\!-\!x)s)\big|^2}\,
\frac{R((1\!-\!x)s)\,\epsilon((1\!-\!x)s)}{R(s)\,\epsilon(s)},
\end{equation}
where $\mathcal{F}(s,x)$ and $\Pi$ are the radiative correction
kernel~\cite{KF:1985} and the  vacuum polarization operator,
respectively. The variable $x$ is a fraction of $s$ lost as a result of
initial-state radiation.

The calculation of the radiation correction is presented in detail in
Section~\ref{sec:radeff}.
\subsection{Monte Carlo simulation}\label{sec:MC}
The KEDR simulation program is based on the GEANT package,
version 3.21~\cite{GEANT:Tool}. 

Single-photon annihilation to hadrons was simulated
using the LUARLW \cite{LUARLW:2001} generator, which 
was employed by the BES collaboration for the high-precision measurement 
of the $R$ value \cite{BES:R2009}.
As an alternative, to simulate $\text{uds}$ continuum we employed the
JETSET~7.4 code \cite{JETSET,PYTHIA64} with the parameters tuned at
energy points 1, 2, 4, 9, 10 and 13. 

Bhabha events required for the precise luminosity determination 
and  $\mu^{+}\mu^{-}$ background process were simulated using the MCGPJ generator~\cite{MCGPJ}.
To simulate two-photon processes $e^{+}e^{-}\to e^{+}e^{-} X$, we employed  
the generators described in Refs. \cite{BERENDS:EEEE,BERENDS:EEMM,KEDR:EEX}.

\begin{figure*}[!ht]
\begin{center}
  \includegraphics[width=0.432\textwidth,height=0.177\textheight]{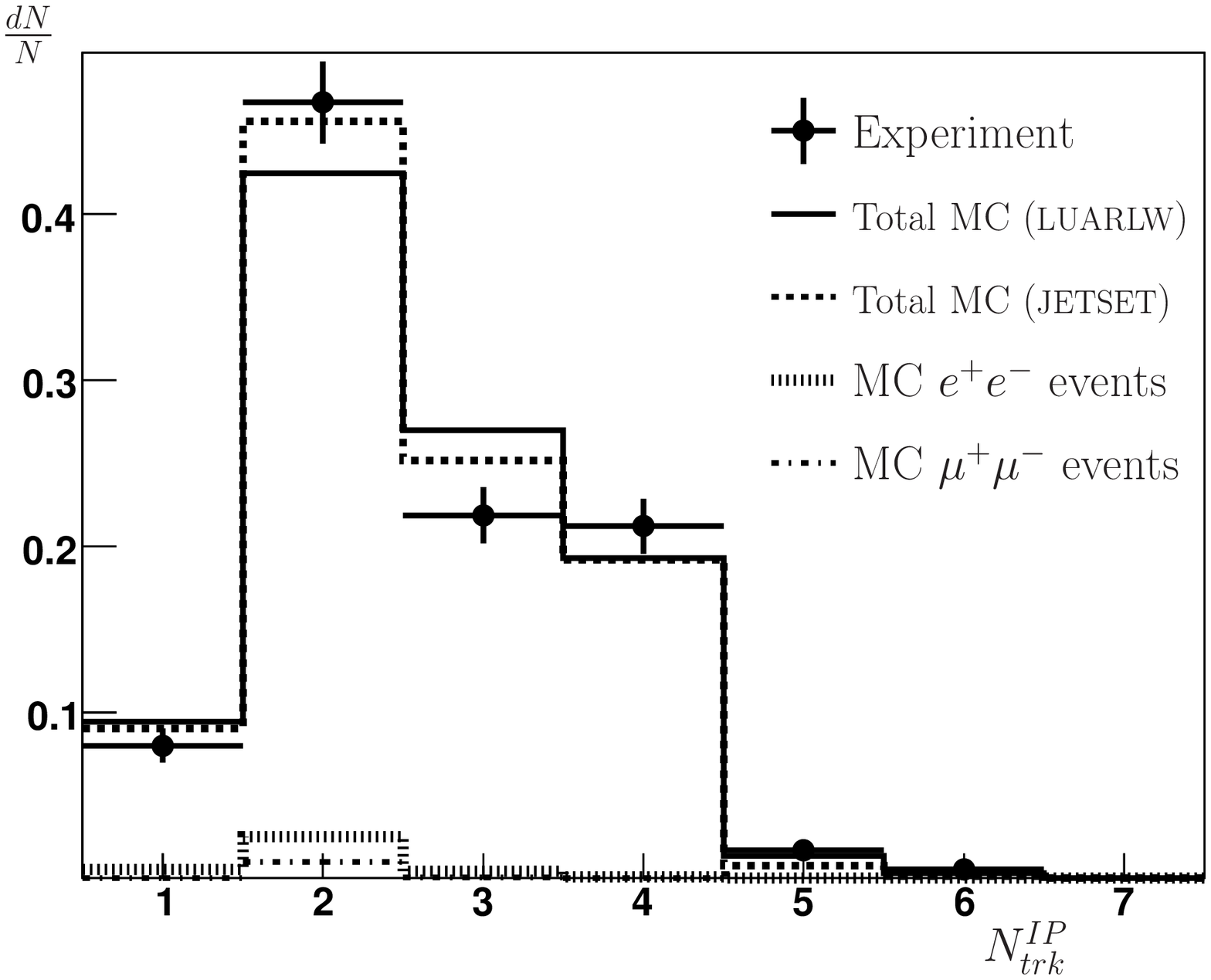}
  \includegraphics[width=0.407\textwidth,height=0.175\textheight]{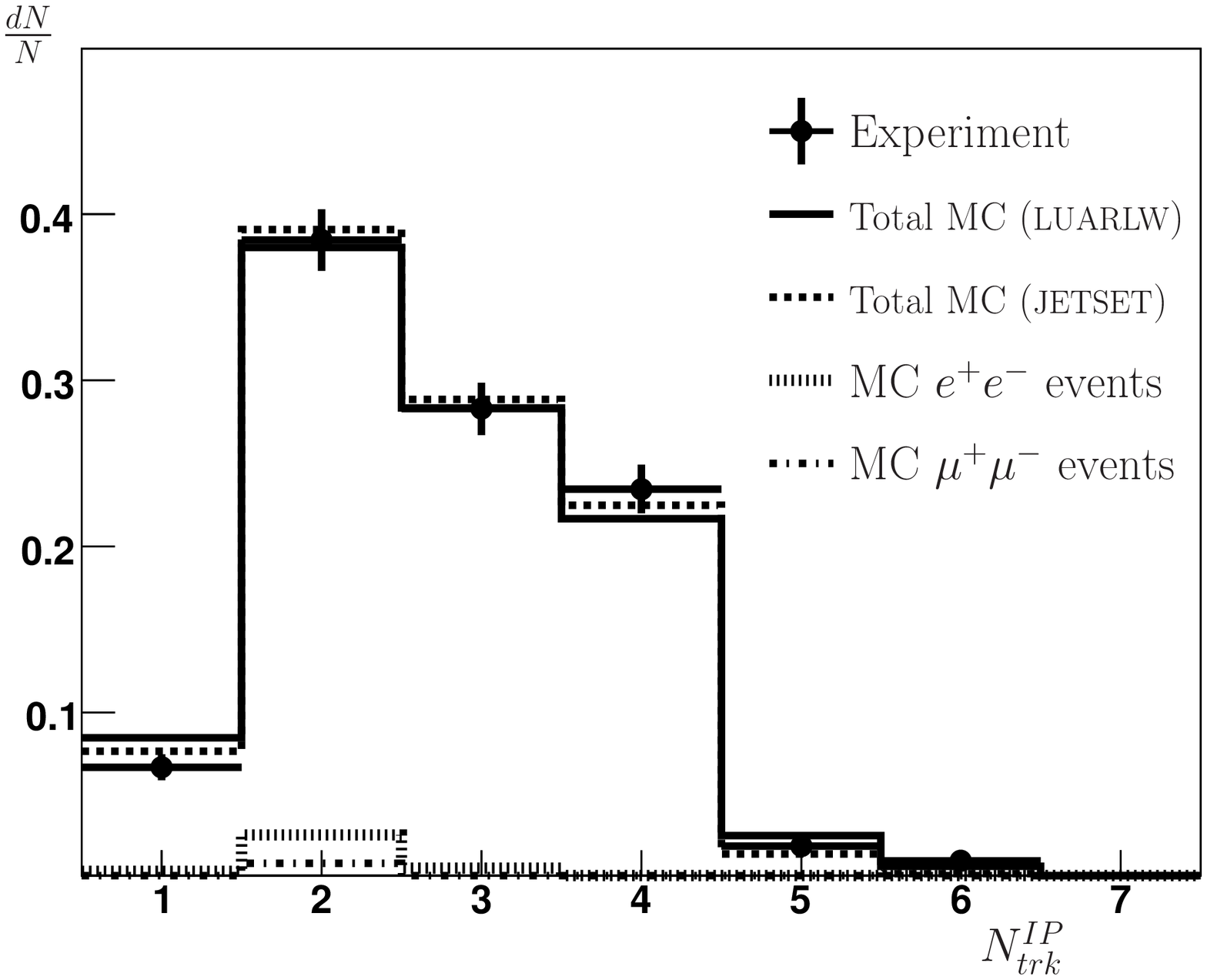}\\
  \includegraphics[width=0.428\textwidth,height=0.179\textheight]{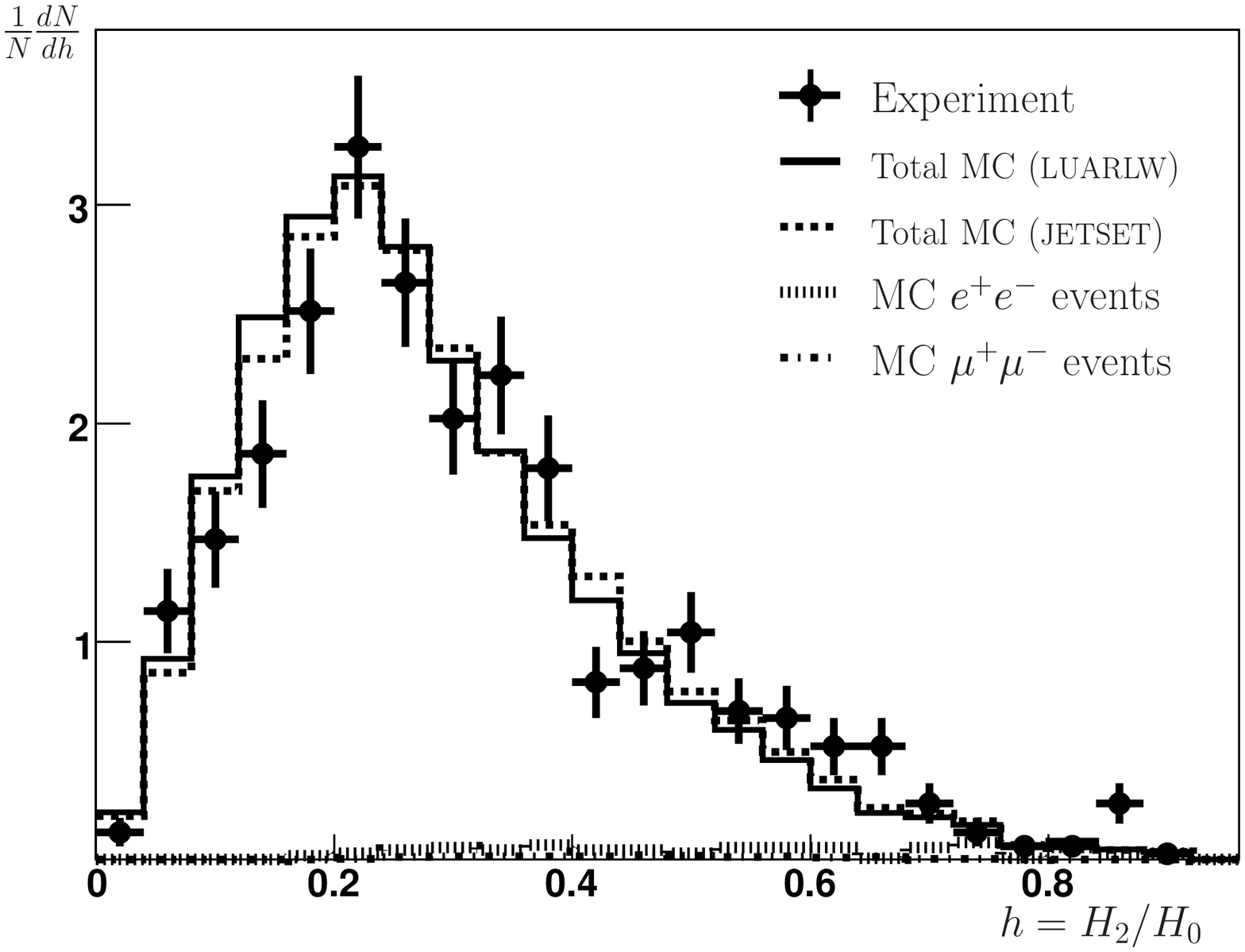}
  \includegraphics[width=0.407\textwidth,height=0.179\textheight]{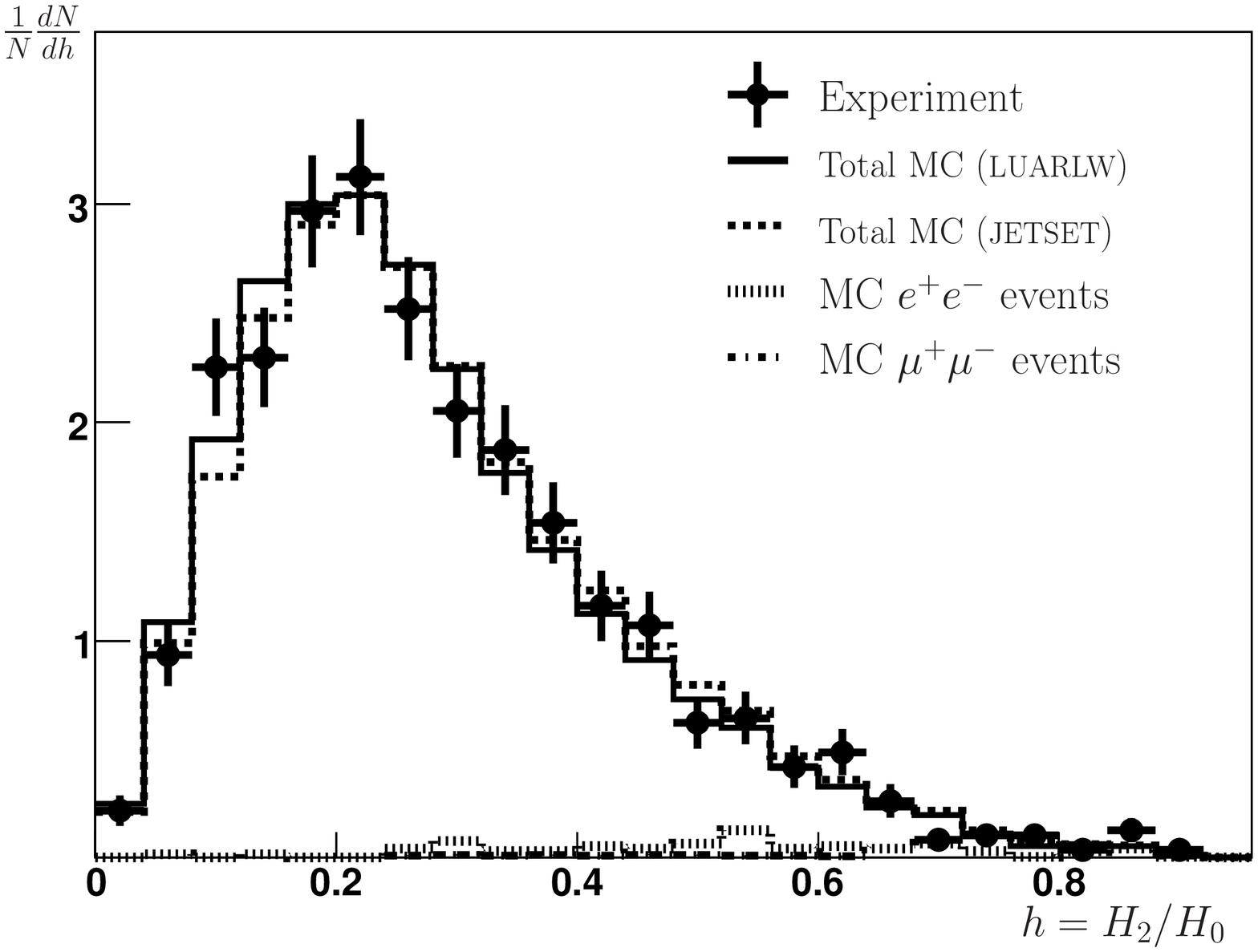}\\
\vspace*{0.2cm}
{
\hspace*{-0.25cm}
  \includegraphics[width=0.415\textwidth,height=0.172\textheight]{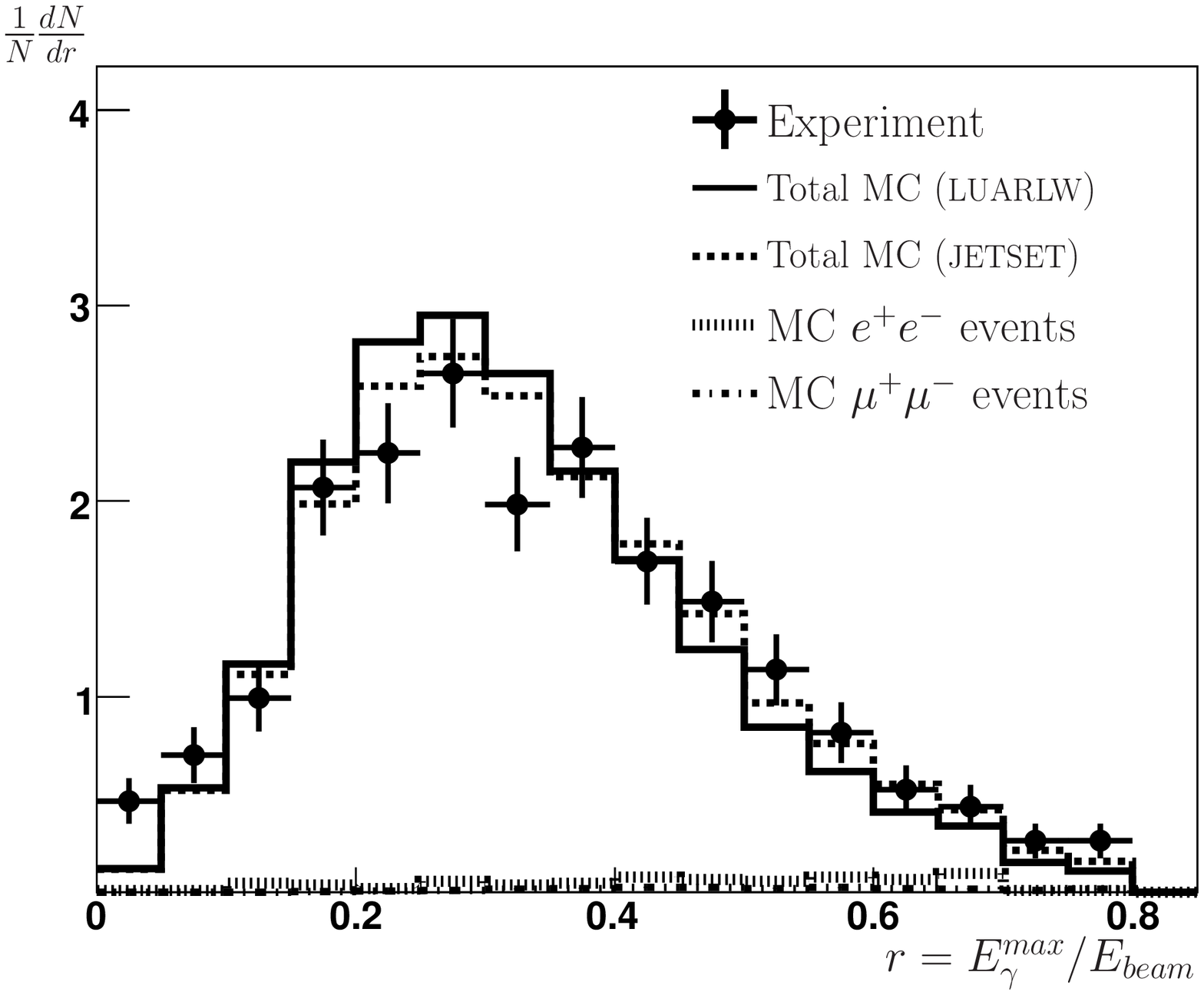}
\hspace*{0.25cm}
  \includegraphics[width=0.415\textwidth,height=0.172\textheight]{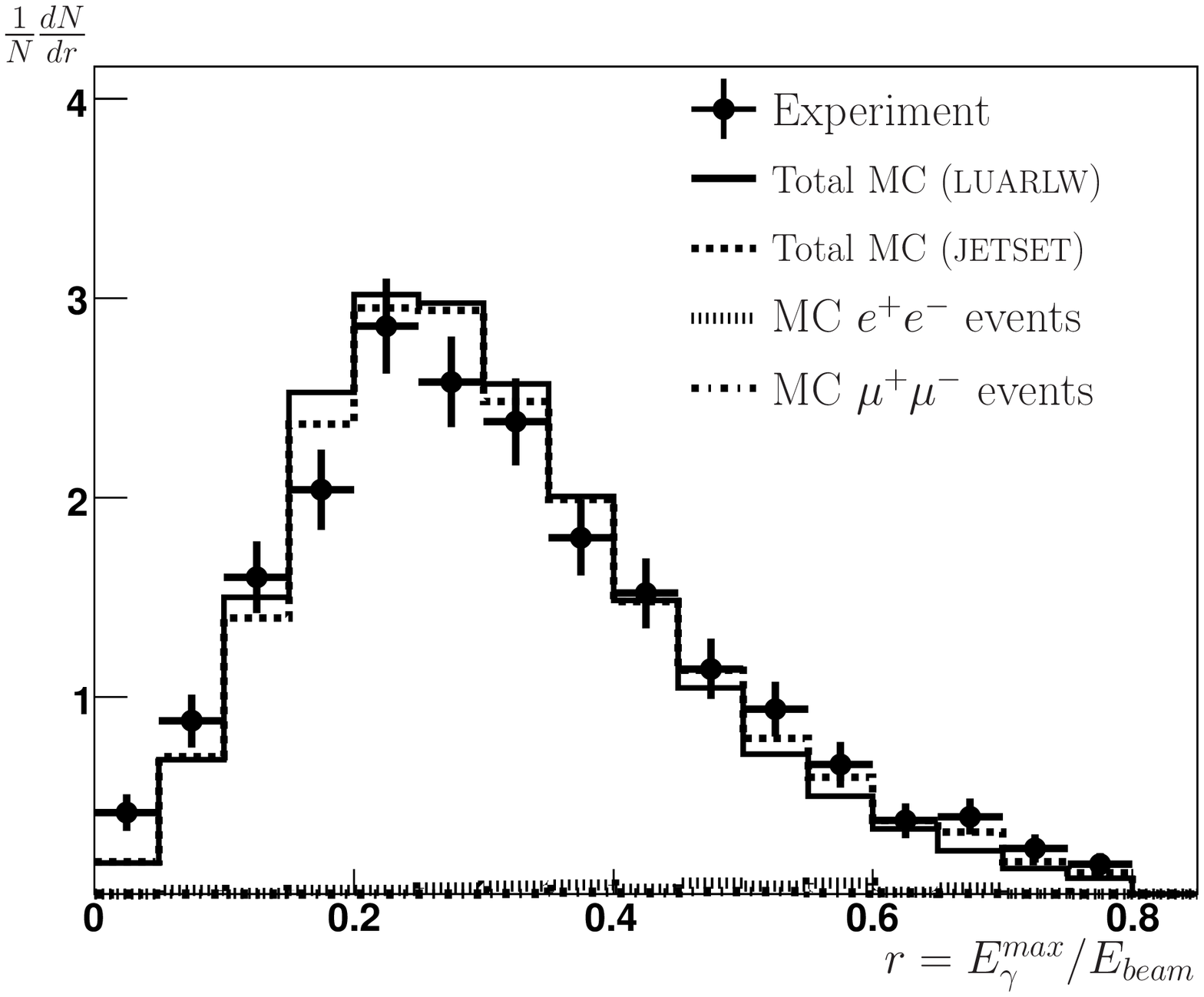} \\
}

{\hspace*{-0.6cm} \includegraphics[width=0.424\textwidth,height=0.186\textheight]{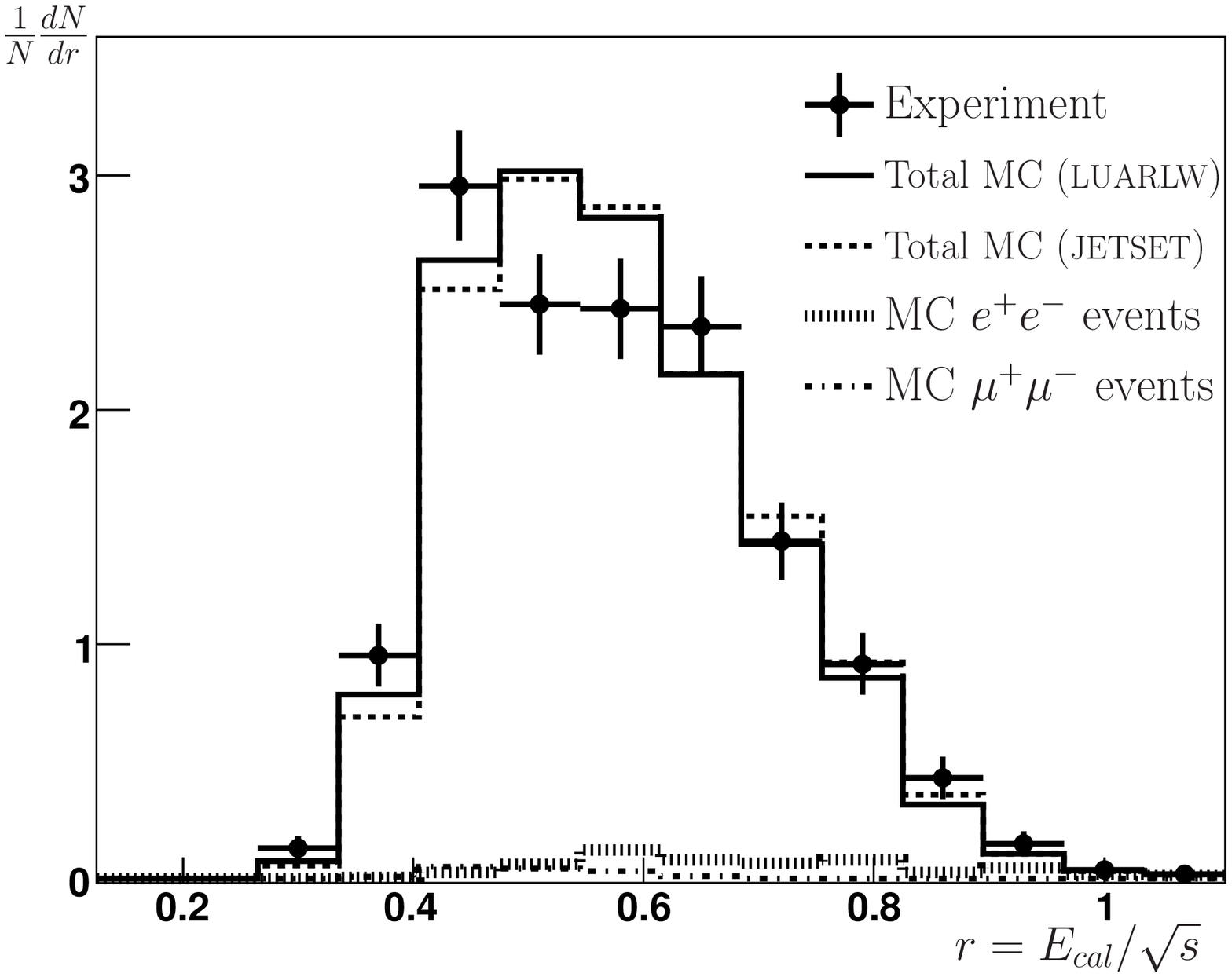}
  \includegraphics[width=0.422\textwidth,height=0.1775\textheight]{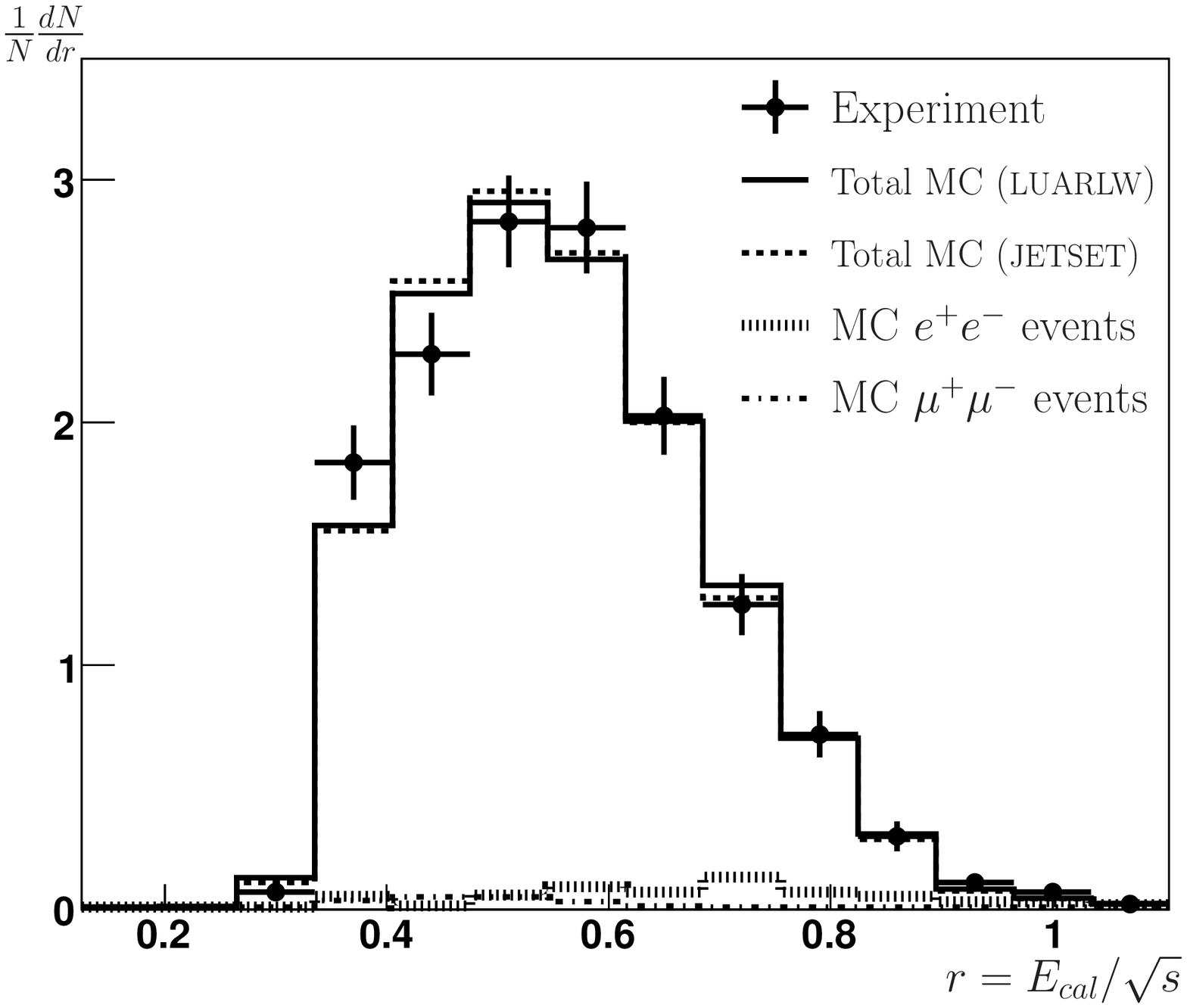}
}
{\hspace*{-0.8cm}\vspace*{0.1cm}
 \includegraphics[width=0.43\textwidth,height=0.178\textheight]{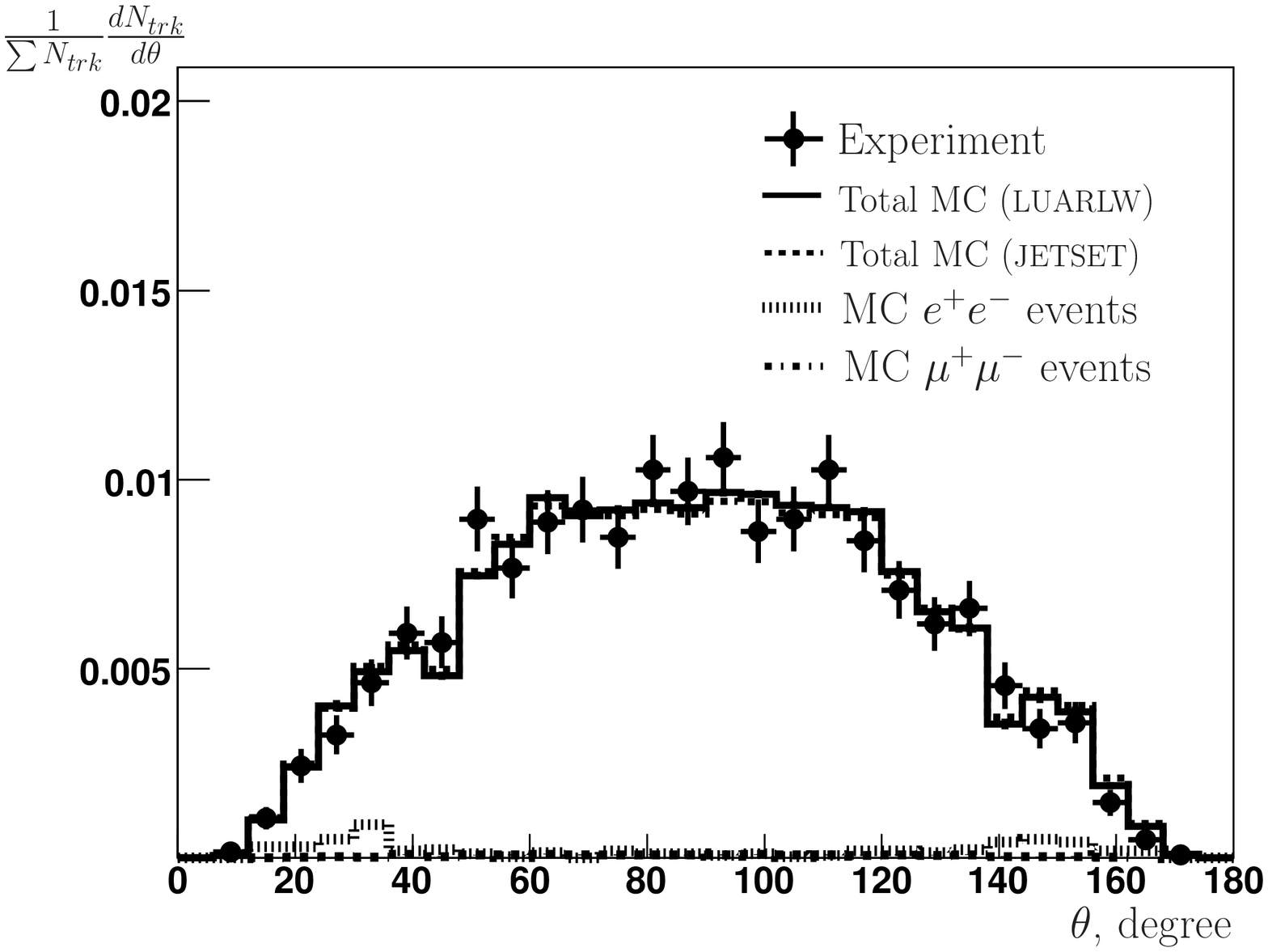}
{\hspace*{-0.7cm}
 \includegraphics[width=0.46\textwidth,height=0.178\textheight]{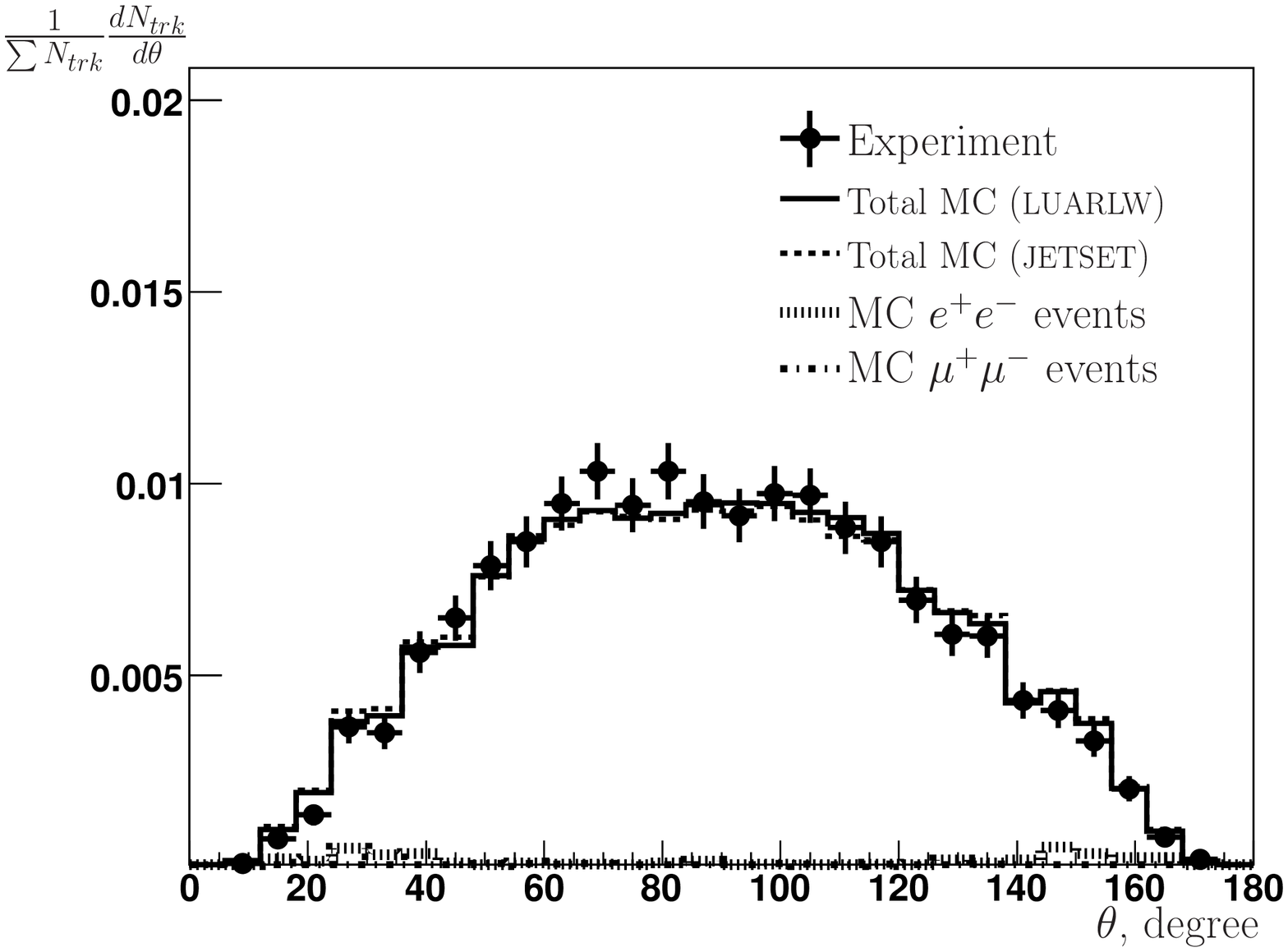}}}

\end{center}
\caption{{ 
 Properties of hadronic events produced in $\text{uds}$ continuum at
 1.94 GeV (left) and 2.14 GeV (right). Here, $N$ is the number of events,
 $H_2$ and $H_0$ are Fox-Wolfram moments \cite{Fox:Wolfram},
 $E_{\gamma}^{\text{max}}$ is  energy of the most energetic photon, 
 $E_{\text{cal}}$ is  energy deposited in the calorimeter, 
 $\theta$ is polar angle, $N_{trk}$ is the number of tracks in event.
 The experimental distribution and two variants of MC simulation based on 
LUARLW and JETSET are plotted. All distributions are normalized to unity. 
Contributions of leptonic pair production are also presented.
}}
\label{simdist_fig1}
\end{figure*}

The results are presented in Fig.~\ref{simdist_fig1}, where the most important 
event characteristics obtained in the experiment are compared with those 
in simulation. Reasonable agreement is observed at all energies. 

It is worth noting that tuning of the JETSET parameters for individual energy points
allows one to reach better agreement between data and Monte Carlo than that
achieved with the LUARLW generator in which the primary event
multiplicity is a function of energy fixed beforehand and the distribution function in it is energy independent. However, at energies below 3~GeV
tuning requires large efforts which are not adequate for the statistically
limited analysis. For this reason, unlike our work~\cite{KEDR:R2016},
it was done at some points only.
The detection efficiencies obtained with the two event generators
are listed in Table \ref{tab:def}.
The systematic uncertainties related to event simulation are discussed
below in Section~\ref{sec:mchadrerr}.
\subsection{Event selection and detection efficiencies}\protect\label{subsec:mhsel}
During the offline analysis, both experimental and simulated events
pass the software event filter. 
This procedure allows us to reduce systematic inaccuracy
due to trigger instabilities  and uncertainties in the hardware thresholds.
The software  filter recomputes  the PT and ST decisions 
with stringent conditions using a digitized response of the detector subsystems.

To suppress the machine background to an acceptable level, the 
following PT conditions were used by OR:
\begin{itemize}\itemsep=-2pt
\item signals from  two or more  non-adjacent scintillation counters\,,
\item signal from the LKr calorimeter\,, 
\item coincidence of the signals from two CsI endcaps.
\end{itemize}
Signals from two particles with the angular separation $\gtrsim\!20^\circ$
should satisfy numerous ST conditions.
The MC simulation yields the trigger efficiency of about 0.94 for 
continuum $\text{uds}$ production.

\renewcommand{\arraystretch}{1.2}
\begin{table}[ht!]
\caption{ \label{tab:criteria} {  Selection criteria for 
hadronic events which were used by AND.}}
\begin{center}
\begin{tabular}{l|l}                                                    

Variable                                        &   Allowed range \\ \hline   
$N^{\text{IP}}_{\text{trk}}$                  &    $\geq 1$\\ \hline        
$E_{\text{obs}}$                                &    $>1.4~\text{GeV}$ ($>1.3~\text{GeV}$ if $E_{beam}<1.05~\text{GeV}$)  \\ \hline
$E_{\text{obs}}-E_{\gamma}^{\text{max}}$        &    $>1.2~\text{GeV}$ ($>1.1~\text{GeV}$ if $E_{beam}<1.05~\text{GeV}$)  \\ \hline
$E_{\gamma}^{\text{max}}/E_{\text{beam}}$       &    $<0.8$ \\ \hline
$E_{\text{cal}}$                   &    $>0.55~\text{GeV}$ \\ \hline
$H_2/H_0$                                       &    $<0.9$ \\ \hline
$|P_{\text{z}}^{\text{miss}}/E_{\text{obs}}|$   &    $<0.6$ \\ \hline
$E_{\text{LKr}}/E_{\text{cal}}$    &    $>0.15$ \\ \hline
$|Z_{\text{vertex}}|$                           &$<15.0~\text{cm}$\\ \hline         
\multicolumn{2}{c}{ $N_{\text{particles}} \geq 3~\text{or}~\tilde{N}^{\text{IP}}_{\text{trk}} \geq 2 $}\\          
\end{tabular}
\end{center}
\end{table}

\renewcommand{\arraystretch}{1.1}
\begin{table}[t!]
\caption{ \label{tab:def} {Detection efficiency for the $\text{uds}$ continuum} in $\%$ (statistical errors only).}
\begin{center}
\begin{tabular}{cccc}    
Point       & $\epsilon_{LUARLW}$      &  $\epsilon_{JETSET}    $   &   $\delta \epsilon/\epsilon$    \\ \hline               
1           & $42.2 \pm 0.1 $   & $45.0 \pm 0.1$ &  $-6.6\pm 0.3$   \\ \hline
2           & $47.2 \pm 0.1 $   & $46.0 \pm 0.1$ &  $-2.5\pm 0.3$   \\ \hline 
3           & $53.4 \pm 0.1 $   & &     \\ \hline                                
4           & $52.5 \pm 0.1 $   & $51.3\pm0.1$ & $-1.2\pm 0.3$   \\\hline  
5           & $57.0 \pm 0.1 $   & &      \\ \hline 
6           & $61.6 \pm 0.1 $   & &     \\ \hline  
7           & $64.3 \pm 0.1 $   & &     \\ \hline
8           & $66.7 \pm 0.1 $   & &         \\ \hline 
9           & $68.2 \pm 0.1 $   &$68.0 \pm0.1$ & $-0.2 \pm 0.2$      \\ \hline 
10          & $70.3 \pm 0.1$    &$70.6 \pm0.1$ & $+0.4 \pm 0.2$            \\ \hline  
11          & $71.6 \pm 0.1$    & &           \\ \hline  
12          & $73.0 \pm 0.1$    & &           \\ \hline  
13          & $72.4 \pm 0.1$    & $73.2\pm 0.1$ & $+1.1\pm 0.2$         \\         
\end{tabular}
\end{center}
\end{table}
\renewcommand{\arraystretch}{1.1}
Selection criteria for multihadron events are listed in Table~\ref{tab:criteria}, 
and their description is provided below.  
In the Table $N^{IP}_{trk}$ is the number 
of tracks from the interaction 
region defined by conditions \mbox{$\,\rho\!<\!5$}~mm,\, \mbox{$|z_0|\!<\!130$}~mm, 
where $\rho$ is the track impact parameter relative to the beam axis,
and $z_0$ is the coordinate of the closest approach point.
The $\tilde{N}^{IP}_{trk}$ is the number of tracks satisfying the 
conditions above with $E/p$  less than 0.6, where $E/p$ means the ratio of 
the energy deposited in the calorimeter to the measured momentum of the 
charged particle.
The multiplicity $N_{\text{particles}}$ is a sum of the number of charged tracks 
and the number of neutral particles detected in the calorimeters.

The observable energy  $E_{\text{obs}}$ is defined as a sum  
of the neutral cluster energies measured in the electromagnetic calorimeter  and
charged particle energies computed from the track momenta assuming pion masses.
The observable energy cut and limitation on the ratio of the energy of 
the most energetic photon 
to the beam energy $E_{\gamma}^{\text{max}}/E_{\text{beam}}$ suppress production 
of \\ hadronic events at low center-of-mass energies 
through initial-state radiation  and thus reduce the uncertainty of 
radiative corrections.   
The total calorimeter energy $E_{\text{cal}} $ is defined as a sum 
of the energies of all clusters 
in the electromagnetic calorimeter.  
The cut on it suppresses the machine background.
The cut on the ratio of Fox-Wolfram moments $H_{2}/H_{0}$  is efficient 
for suppression of the $\ee\!\to\!\ee\gamma\,$
background, that of cosmic rays and some kinds of the machine background.
The background from two-photon and beam-gas events is suppressed by the cut 
on the ratio $|P_{\text{z}}^{\text{miss}}/E_{\text{obs}}|$,  where $P_{\text{z}}^{\text{miss}}$ 
is the $\text{z}$ component of missing momentum.
The background from beam-gas events was also suppressed by the cut 
on the ratio  $E_{\text{LKr}}/E_{\text{cal}}$ 
of the energy deposited in the LKr calorimeter and total calorimeter energy.
The event vertex position  $Z_{\text{vertex}}$ is the weighted average
of the $z_0$'s of
the charged tracks. The cut on the $|Z_{\text{vertex}}|$ suppresses
background due to beam-gas, beam-wall and cosmic rays.

The muon system veto was required to reject cosmic rays background 
in the cases when more than two tracks did not cross 
the interaction region or the event arrival time determined by TOF
relative to the bunch crossing was less than -7 ns or larger than 12 ns.

Compared with our previous work \cite{KEDR:R2016}, we have introduced
an additional condition on the difference of the observable energy and
the energy of  the most energetic photon that reduces the uncertainty of 
radiative corrections. At the same time, some of selection conditions
were relaxed to increase the detection efficiency below 2.5~GeV. 

The detection efficiency for hadronic events corresponding
to the selection criteria described above is presented in
Table~\ref{tab:def} for thirteen data points at which the $R$ ratio
was measured. For six energy points it was determined using two
versions of the event simulation.

\subsection{Luminosity determination}\protect\label{subsec:lum}
The integrated luminosity at each point was
determined using Bhabha events detected in the LKr calorimeter in the
polar angle range $44^{\circ}\!<\!\theta\!<\!136^{\circ}$. 

The criteria for \ee event selection are listed below:
\begin{itemize}\itemsep=-2pt
\item  two clusters, each with the energy above $20\%$ of the beam 
energy and the angle between them exceeding $162^{\circ}$,
\item the total energy of these two clusters exceeds the beam energy,
\item the calorimeter energy not associated  with
      these two clusters does not exceed 30$\%$ of the total.
\end{itemize}
The tracking system was used only to reject the background from 
$\ee\!\to\!\gamma\gamma$ and $\ee\!\to\!\text{\it hadrons}$.
\subsection{Physical background}\protect\label{sec:physbackground}

To determine $R$ values, we took into account the physical background 
contributions from the QED processes $\ee\to\ee$ and $\ee\to\mumu$
which are summarized in Table \ref{tab:effbckg}.

The contributions of two-photon interactions were studied with a
simulation of $\ee\to\ee X$~events. We  found that the contribution of two-photon events to the continuum cross section grows from $0.1\%$ 
at 1.84 GeV to $0.3\%$ at 3.05 GeV.
The estimated uncertainty in the $R$ value due to this contribution
is less than $0.2\%$.

\begin{table}[ht]
\caption{\label{tab:effbckg} {\normalsize The contribution of the physical background to the observed cross section in \%.}}
\begin{center}
\begin{tabular}{lcc}        
Point           &\multicolumn{2}{c}{ Process}  \\ 
                & $\ee$           &   $\mumu$          \\ \hline
 1              & $6.07 \pm  0.56$&  $1.08 \pm 0.04$     \\ \hline
 2              & $4.13 \pm  0.45$&  $1.06 \pm 0.03$      \\ \hline
 3              & $3.70 \pm  0.39$&  $0.99 \pm 0.03$       \\ \hline
 4              & $3.81 \pm  0.39$&  $1.00 \pm 0.03$                \\ \hline
 5              & $4.93 \pm  0.43$&  $0.96 \pm 0.03$   \\ \hline
 6              & $4.40 \pm  0.39$&  $1.02 \pm 0.03$   \\ \hline
 7              & $3.30 \pm  0.34$&  $0.87 \pm 0.03$  \\ \hline
 8              & $4.22 \pm  0.37$&  $0.85 \pm 0.03$      \\ \hline
 9              & $4.74 \pm  0.39$&  $0.81 \pm 0.03$    \\ \hline
 10             & $4.12 \pm  0.35$&  $0.80 \pm 0.03$        \\ \hline
 11             & $4.74 \pm  0.38$&  $0.82 \pm 0.03$          \\ \hline
 12             & $5.07 \pm  0.38$&  $0.82 \pm 0.03$     \\ \hline
 13             & $5.88 \pm  0.41$&  $0.83 \pm 0.03$      \\ 
\end{tabular}
\end{center}
\end{table}

\subsection{Correction for machine background}\protect\label{sec:background}
To estimate the contribution of residual machine background to 
the observed cross section, 
we use runs with separated $e^{+}$ and $e^{-}$ bunches.

The number of events that passed selection criteria in the runs with 
separated bunches was recalculated to the number of expected background events 
under the assumption that the background rate is proportional to the beam
current and the measured vacuum pressure.
As an alternative, we also performed analysis assuming  that the background
rate is proportional to the current only.
The difference between the numbers of background events obtained with
the two assumptions was considered as an uncertainty estimate at given
energy point.

The background values and their uncertainties at each energy point 
are presented in Table~\ref{tab:background}.
\begin{table}[h!]
\caption{{\label{tab:background} The residual machine 
background in $\%$ of the observed cross section}}
\begin{center}
\begin{tabular}{cc|cc}
Point   &   Background, \% & Point  &    Background, \%\\\hline        
    1   &   $1.4 \pm 0.5  \pm 0.4$& 8&  $0.4 \pm 0.4  \pm 0.2$ \\ \hline    
    2   &   $1.4 \pm 0.5  \pm 0.2$& 9 &  $1.2 \pm 0.6  \pm 0.2$ \\ \hline    
    3   &   $1.2 \pm 0.4  \pm 0.2$& 10 &   $1.8 \pm 0.7  \pm 0.3$ \\ \hline    
    4   &   $1.6 \pm 0.7  \pm 0.2$& 11& $1.2 \pm 0.4  \pm 0.2$ \\ \hline    
    5   &   $2.1 \pm 0.8  \pm 0.2$& 12&   $2.1 \pm 0.9  \pm 0.2$\\ \hline    
    6   &   $1.3 \pm 0.6  \pm 0.2$& 13&  $1.3 \pm 0.5  \pm 0.2$\\ \hline    
    7   &   $0.6 \pm 0.6  \pm 0.6$&   &   \\     
\end{tabular}
\end{center}
\end{table}

\subsection{Radiative correction}\protect\label{sec:radeff}
The radiative correction factor was determined according to
Eq.~\eqref{eq:RadDelta} using the compilation of the vacuum polarization data by the CMD-2 group
\cite{Actis:2010gg} and the relation between 
$R(s)$ and the hadronic part of the vacuum polarization $\Pi_{\text{hadr}}(s)$:
\begin{equation}
R(s)=-\frac{3}{\alpha} \Imag \Pi_{\text{hadr}}(s).
\end{equation}

\begin{figure}[!ht]
\includegraphics[width=0.48\textwidth]{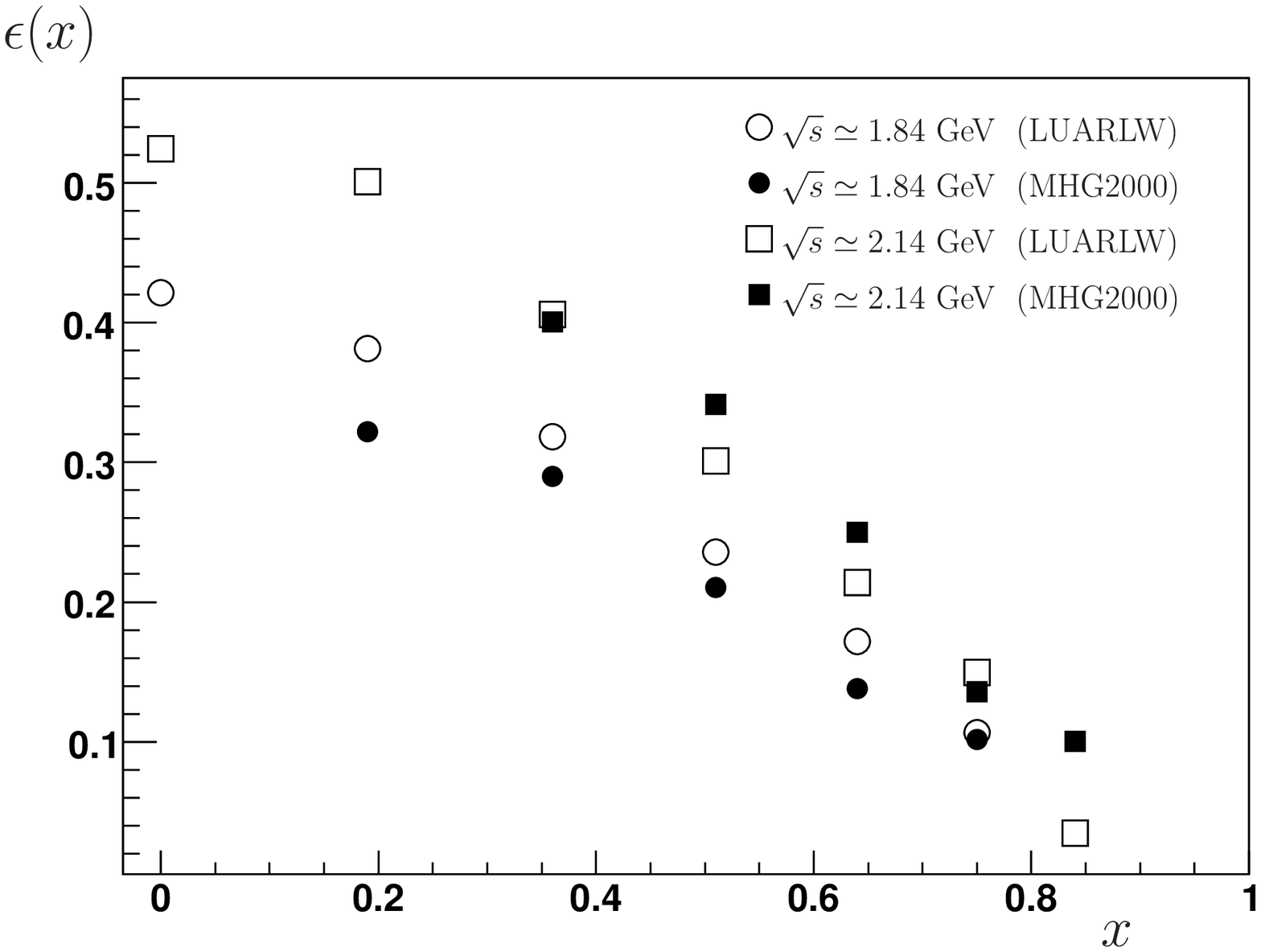}
\caption{{\label{fig:radeff} Hadronic detection efficiency versus
    variable $x$ of Eq.~\eqref{eq:RadDelta} at 1.84 and 2.14 GeV.
}}
\end{figure}

For each energy, the dependence of the detection efficiency on the energy radiated  in the initial state was evaluated with the $\text{LUARLW}$  and 
the MHG2000 generator, the latter developed by the CMD-3 
collaboration \cite{CMD3MC1,CMD3MC2}. 
We apply the MHG2000 generator to simulate hadronic events below 1.84 GeV.
This generator simulates about 30 various exclusive modes and
approximately reproduces a real picture of $e^+e^-\to hadrons$ below 2 GeV. 
The $x$ dependencies of the detection efficiencies obtained with  the
$\text{LUARLW}$ and MHG2000 generators  for some energies 
are shown in Fig.~\ref{fig:radeff}. 

Table \ref{tab:delta} contains values of the radiative correction and their 
systematic uncertainties which are discussed in Sec.~\ref{sec:raderr}.
\begin{table}[th!]
\caption{ \label{tab:delta}{Radiative correction factor $1+\delta$}}
\begin{center}
\begin{tabular}{cc|cc}    
 Point         &     $1+\delta$       &  Point    &   $1+\delta$ \\ \hline
1              & $1.0423 \pm 0.0208$ & 8   &  $1.0739 \pm 0.0054 $ \\ \hline
2              & $1.0429 \pm 0.0156$ & 9   &  $1.0796 \pm 0.0054 $ \\ \hline
3              & $1.0515 \pm 0.0126$ & 10  &  $1.0809 \pm 0.0054 $ \\ \hline
4              & $1.0634 \pm 0.0106$ & 11  &  $1.0823 \pm 0.0054 $    \\ \hline
5              & $1.0645 \pm 0.0096$ & 12  &  $1.0774 \pm 0.0054 $    \\ \hline
6              & $1.0664 \pm 0.0075$ & 13  &  $1.0584 \pm 0.0053 $ \\ \hline
7              & $1.0684 \pm 0.0064$ &     &   \\ 
\end{tabular}
\end{center}
\end{table}

\section{Systematic uncertainties and results}\protect\label{sec:systerr}
\subsection{Systematic uncertainty of absolute luminosity
  determination}\protect\label{sec:lumerr}
A summary of  systematic uncertainties in the
absolute luminosity determination with the LKr calorimeter is
given in Table~\ref{tab:lumerr}.

\begin{table}[h]
\caption{{ \label{tab:lumerr} Systematic uncertainties of the luminosity determination.}}
\begin{center}
\begin{tabular}{lc}
Source &  Uncertainty, $\%$   \\ \hline                            
Calorimeter response          & 0.7  \\  \hline
Calorimeter alignment         & 0.2  \\  \hline
Polar angle resolution        & 0.2  \\  \hline
Cross section calculation     & 0.5  \\  \hline 
Background                    & 0.1  \\  \hline
MC statistics                 & 0.1 \\   \hline
Variation of cuts             & 0.8 \\   \hline 
Sum in quadrature             & 1.2 \\
\end{tabular}
\end{center}
\end{table}

The uncertainty  related to the imperfect simulation of the calorimeter
response was estimated by variation of relevant simulation parameters such as
the geometrical factor controlling sensitivity to the energy loss
fluctuations between calorimeter electrodes, the accuracy of the
electronic channel calibration, etc.
 
The alignment of the LKr calorimeter with respect to the drift chamber   
was done with the help of cosmic tracks reconstructed in the tracking system.
The  direction of the beam line and interaction point were determined
using the primary vertex distribution of multihadron events. 
The luminosity   uncertainty  due to inaccuracy of the alignment  is less 
than $0.2\%$.

The difference in the polar angle resolutions observed in  experiment
and predicted by simulation causes an uncertainty in the luminosity 
measurement because events migrate  into or out of the fiducial volume.

The uncertainty of the theoretical Bhabha cross section
was estimated comparing the results obtained with
the MCGPJ~\cite{MCGPJ} and BHWIDE~\cite{BHWIDEGEN}   event generators.
It agrees with the errors quoted by the authors.

The background to the Bhabha process  from the  reactions 
$\ee \to \mu\mu(\gamma)$ and $\ee \to \gamma\gamma $ 
was estimated using MC simulation. 
It contributes less than 0.2\% to the observed $e^{+}e^{-}$ cross section
for all energy points presented in Table~\ref{tab:epoints}.
We also considered a contribution of residual machine background  
to Bhabha events which is about $0.1\%$.
The residual luminosity uncertainty due to background does not exceed $0.1\%$. 

In order to estimate the effect of other possible sources of
uncertainty, we varied cuts within the fiducial region.
The cuts on the polar angle were varied in a range much larger than the
angular resolution, the variation in the Bhabha event count
reaches $50\%$. 
The cut on the deposited energy was varied in the range of $50-80\%$
of the c.m. energy. 
The variations discussed above correspond to a systematic uncertainty
shown in Table~\ref{tab:lumerr}. 

These effects can occur due to the already considered
sources and statistical fluctuations, nevertheless we add
them in the total uncertainty to obtain conservative error estimates.

\subsection{Uncertainty due to imperfect simulation of continuum}\protect\label{sec:mchadrerr}
The imperfect simulation of the $\text{uds}$ continuum contributes
significantly to the systematic uncertainty in  $R$. 
The maximal deviation of $1.2\%$ is taken from Table \ref{tab:def} as 
the systematic  uncertainty for the energy range 2.14-3.05 GeV.
This estimate is consistent with  our previous result for the  LUARLW
generator uncertainty of $1.3\%$ obtained from continuum simulation above
the $J/\psi$ \cite{KEDR:R2016}. 
Below 2.14 GeV our estimations of this uncertainty are $2.5\%$ 
for points 2 and 3 and  $6.6\%$ for point 1.
These estimations were checked
by the variation of selection criteria described in
Section~\ref{err:deterr}.


The contributions to the detection efficiency uncertainty due to 
imperfect simulation of the $\text{uds}$ continuum 
are summarized in Table~\ref{tab:mcerr}.

\begin{table}[ht]
\caption{{\label{tab:mcerr} Systematic uncertainties of the
detection efficiency due to imperfect simulation of continuum. 
}}
\begin{center}
\begin{tabular}{lccc}                     

Source                & \multicolumn{3}{c}{ Uncertainty, $\%$}  \\ \hline
                       &Point 1 & Points 2-3  & Points 4-13  \\ \hline
$\text{uds}$ simulation&6.6 &  2.5       & 1.2    \\ \hline
MC statistics          &0.2 &  0.2        & 0.2   \\ \hline
Sum in quadrature      &6.6 &  2.5       & 1.2 \\
\end{tabular}
\end{center}
\end{table}

\subsection{Systematic uncertainty of the radiative correction}\protect\label{sec:raderr}
The main sources of systematic uncertainty associated with the 
radiative correction factor at each energy point are summarized 
in Table \ref{tab:raderr}.

To estimate the uncertainty related to the accuracy of the vacuum
polarization operator, we have compared two approximations to it.
The first one was obtained by the CMD-2 group \cite{Actis:2010gg}, the
alternative one was extracted from the BES event generator~\cite{BESGEN}.
The difference between them reaches 1.4\% at the
lowest energy point 1.84 GeV and drops down to $0.3\%$ 
at the energy above 2.2~GeV.

The contribution denoted as $\delta \epsilon(s)$
is related to the uncertainty in the $\epsilon(s)$ dependence and obtained from
the two alternative simulations below 1.84 GeV with the MHG2000 and 
LUARLW generators.  
The obtained difference in the $1+\delta$ value for 
each energy point is assigned as a systematic error due to $\epsilon(s)$ uncertainty.

The contribution $\delta_{\rm calc}$ is related to the interpolation 
uncertainty. It was evaluated comparing the results obtained using the linear
interpolation and the quadratic one.

The estimated uncertainty in the radiative correction due to 
the $R(s)$ uncertainty varies from 0.5\% to 2.0\%  
for the entire  energy range.

\begin{table}[h!]
\caption{{ \label{tab:raderr} Systematic uncertainties of the 
radiative correction.}}
{\small
\begin{center}
\begin{tabular}{cccccc}

           &  \multicolumn{5}{c}{ Uncertainty, $\%$ }     \\ \hline
  Point  &  \multicolumn{4}{c}{Contributions} & Total \\ \hline
   & $\Pi$ approx.  &$\delta \epsilon(s)$ &   $\delta R(s)$& $\delta_{calc}$ &  \\ \hline    
    1   &   1.3 & 1.4&0.6&0.1& 2.0    \\ \hline %
    2   &   1.1 & 1.0&0.4&0.1& 1.5   \\ \hline  %
    3   &   1.0 & 0.6&0.4&0.1& 1.2   \\ \hline  %
    4   &   0.8 & 0.4&0.4&0.1& 1.0   \\ \hline  %
    5   &   0.7 & 0.3&0.4&0.1& 0.9  \\ \hline   %
    6   &   0.6 & 0.3&0.2&0.1& 0.7  \\ \hline   %
    7   &   0.5 & 0.3&0.2&0.1& 0.6   \\ \hline  %
    8   &   0.4 & 0.3&0.2&0.1& 0.5   \\ \hline  %
    9   &   0.3 & 0.3&0.2&0.1& 0.5   \\ \hline  %
    10  &   0.3 & 0.3&0.2&0.1& 0.5   \\ \hline  %
    11  &   0.3 & 0.3&0.2&0.1& 0.5   \\ \hline  %
    12  &   0.3 & 0.3&0.2&0.1& 0.5   \\ \hline   %
    13  &   0.4 & 0.3&0.2&0.1& 0.5   \\   %
\end{tabular}
\end{center}
}
\end{table}

\subsection{Detector-related uncertainties in $R$}\label{err:deterr}
The systematic uncertainties related to the efficiency of the track 
reconstruction were studied using Bhabha events and
low-momentum cosmic tracks, and the appropriate correction was introduced 
in the MC simulation.
The uncertainty of the correction gives the additional systematic
uncertainty of about 0.5\%.

The main source of the trigger efficiency uncertainty is 
that of the calorimeter thresholds in the secondary trigger.
The estimate of about $0.2\%$ was obtained varying the threshold
in the software event filter. The inefficiency of the first level trigger
related to inefficiency of the time-of-flight counters 
is less than $0.2\%$.

The trigger efficiency  and the event selection efficiency depend 
on the calorimeter response to hadrons. 
The uncertainty related to the simulation of nuclear interaction
was estimated by comparison of the efficiencies obtained with the
packages GHEISHA~\cite{Fesefeldt:1985yw} and FLUKA~\cite{Fasso:2005zz}
which are implemented in GEANT~3.21~\cite{GEANT:Tool}. 
The relative difference was about 0.4\%.

To evaluate a systematic uncertainty related to the neutral events
(no tracks in VD and three or more neutral particles)
we add events that met the criteria which are listed in Table \ref{tab:neutralcriteria}.
This selection gives additional $0.7\%$ of hadronic events and 
changes the average R within 0.2\%, 
that serves as our estimate of the systematic 
uncertainty coming from neutral events.
\begin{table}[ht!]
\caption{\label{tab:neutralcriteria} {Selection criteria for neutral events which were used by AND.}}
\begin{center}
\begin{tabular}{l|l}                                                    
Variable                                        &   Allowed range \\ \hline 
\multicolumn{2}{c}{$N_{\text{particles}}^{neutral} \geq 3$ and no tracks in VD}  \\ \hline 
$E_{\text{cal}}$                                & $>E_{\text{beam}}$   \\ \hline
$E_{\text{cal}}-E_{\gamma}^{\text{max}}$        & $>1.2~\text{GeV}$ ($>1.1~\text{GeV}$ if $E_{beam}<1.05~\text{GeV}$)  \\ \hline
$H_2/H_0$                                       & $<0.9$ \\ \hline
$E_{\text{LKr}}/E_{\text{cal}}$    & $>0.5$ \\ \hline
\end{tabular}
\end{center}
\end{table}

The effect of other possible sources of the detector-related uncertainty 
was evaluated by varying the event selection cuts that are presented in 
Table~\ref{tab:criteriavar}. 
All observed $R$ variations were
 smaller than their statistical errors and can originate from 
the already considered sources of uncertainties 
or the statistical 
fluctuations.
Nevertheless, keeping the conservative estimate,  we included 
them in the total uncertainty.
\begin{table}[h!]
\caption{\label{tab:criteriavar} {$R$ uncertainty due to variation of the selection criteria for hadronic events.}}
\begin{center}
\begin{tabular}{llc}      

Variable                                       &   Range variation & $R$ variation in \% \\ \hline   
$E_{\text{obs}}$                               &    $> 1.3  \div 1.7~\text{GeV}$ & 0.3 \\ \hline
$E_{\text{obs}}-E_{\gamma}^{\text{max}}$       &    $> 1.1  \div 1.4~\text{GeV}$ & 0.3    \\ \hline
$E_{\gamma}^{\text{max}}/E_{\text{beam}}$      &    $< 0.6  \div  0.9$           & 0.3 \\ \hline
$E_{\text{cal}}$                  &    $> 0.5  \div  0.8~\text{GeV}$ & 0.2 \\ \hline
$ H_2/H_0$                                     &    $< 0.75  \div  0.92$ & 0.3\\ \hline
$|P_{\text{z}}^{\text{miss}}/E_{\text{obs}}|$  &    $< 0.6  \div  0.8$  & 0.2     \\ \hline
$E_{\text{LKr}}/E_{\text{cal}}$   &    $> 0.15 \div  0.25$& 0.1\\ \hline
$|Z_{\text{vertex}}|$                          &    $< 12.0\div25.0~\text{cm}$&  0.2 \\  \hline
\multicolumn{2}{c}{Sum in quadrature}          &      0.7   \\                             
\end{tabular}
\end{center}
\end{table}
\subsection{Energy determination uncertainty}
During data collection at given energy point, the c.m. energy
uncertainty was about~2~MeV. Meanwhile,
the detection efficiency varied from 0.42 to 0.73 in the 
energy range of the experiment. 
That leads to inaccuracy of the detection efficiency 
determination. Using linear efficiency interpolation between energy 
points, we estimated the contribution of the energy determination
uncertainty to the $R$ systematic error. It is about 0.1\% for the entire  energy range.
 
\subsection{Results}\protect\label{sec:errsummary}
A summary of the systematic uncertainties affecting the 
measurement of $R$ is presented  in Table~\ref{tab:rerr}.
\renewcommand{\arraystretch}{1.}
\begin{table*}[th!]
\caption{{ \label{tab:rerr} $R$ systematic uncertainties (in $\%$) 
assigned to each energy point.}}
\begin{center}
\begin{tabular}{cccccccc}

                           & Point 1  & Point 2  &   Point 3   & Point 4  &   Point 5 & Point 6  & Point 7          \\ \hline    
Luminosity                 &   1.2     &  1.2     &     1.2     & 1.2   &   1.2  & 1.2   & 1.2          \\ \hline     
Radiative correction       &   2.0     &  1.5   &     1.2     & 1.0  &   0.9  & 0.7    & 0.6       \\ \hline    
Continuum simulation       &   6.6     &  2.5   &     2.5     & 1.2   &   1.2  & 1.2    & 1.2         \\ \hline    
Track reconstruction       &   0.5     &  0.5   &      0.5    & 0.5   & 0.5    & 0.5   & 0.5     \\ \hline
$l^+l^-$  contribution     &   0.6     &  0.5   &     0.4     & 0.4   &   0.4  & 0.4   & 0.3          \\ \hline     
$e^+e^-X$  contribution    &   0.2     &  0.2    &     0.2    & 0.2   &   0.2   & 0.2    & 0.2          \\ \hline     
Trigger efficiency         &   0.3     &  0.3   &     0.3    & 0.3   &   0.3  & 0.3    & 0.3          \\ \hline    
Nuclear interaction        &   0.4     &  0.4    &     0.4    & 0.4   &   0.4   & 0.4    & 0.4          \\ \hline   
Neutral events             &   0.2     &  0.2    &     0.2    & 0.2   &   0.2   & 0.2    & 0.2          \\ \hline                  
Cuts variation             &   0.7     &  0.7    &     0.7    & 0.7   &   0.7   & 0.7    & 0.7          \\ \hline    
Machine background         &   0.6     &  0.5    &     0.4    & 0.7   &   0.8   & 0.6    & 0.8         \\  \hline   
Energy determination       &   0.1     &  0.1     &    0.1    & 0.1   &   0.1     & 0.1    &  0.1         \\  \hline   
                          
Sum in quadrature           &   7.1   &  3.4       &    3.2      & 2.4  &   2.4    &   2.2       &  2.2       \\

                           & Point 8  & Point 9   &   Point 10  & Point 11  &   Point 12 & Point 13  &           \\ \hline    
Luminosity                 &   1.2     &  1.2     &     1.2       & 1.2     &   1.2      & 1.2     &           \\ \hline     
Radiative correction       &   0.5      &  0.5    &     0.5       & 0.5     &   0.5      & 0.5     &         \\ \hline    
Continuum simulation       &   1.2      &  1.2    &     1.2       & 1.2     &   1.2      & 1.2     &          \\ \hline
Track reconstruction       &   0.5     &  0.5     &     0.5       & 0.5     &   0.5      & 0.5    &  \\ \hline
     
$l^+l^-$  contribution     &   0.4      &  0.4    &     0.4       & 0.4     &   0.4      & 0.4    &           \\ \hline     
$e^+e^-X$  contribution    &   0.2      &  0.2    &     0.2       & 0.2     &   0.2      & 0.2    &           \\ \hline     
Trigger efficiency         &   0.3      &  0.3    &     0.3       & 0.3     &   0.3      & 0.3    &           \\ \hline    
Nuclear interaction        &   0.4      &  0.4    &     0.4       & 0.4     &   0.4      & 0.4    &           \\ \hline  
Neutral events             &   0.2      &  0.2    &     0.2       & 0.2     &   0.2      & 0.2    &           \\ \hline                  
Cuts variation             &   0.7      &  0.7    &     0.7       &0.7      &   0.7      & 0.7    &     \\ \hline    
Machine background         &   0.4      &  0.6    &     0.8       &0.4      &   0.9      & 0.5    &          \\  \hline   
Energy determination       &   0.1      &  0.1    &     0.1       & 0.1     &   0.1      & 0.1    &          \\    \hline

Sum in quadrature          &   2.1      &  2.2    &     2.2       & 2.1 &   2.3     &   2.1        &      \\ 

\end{tabular}
\end{center}
\end{table*}

Note that the contribution of the $J/\psi$ resonance to the  absolute $R(s)$ 
value
is not completely negligible for the upper point of the energy scan and 
amounts to $6\cdot10^{-3}$.
This contribution was found  analytically
using "bare"  parameters of the resonances, which were calculated
based on the PDG data~\cite{PDG:2014}.  

\begin{figure*}[!]
\begin{center}
\includegraphics[width=0.5\textwidth]{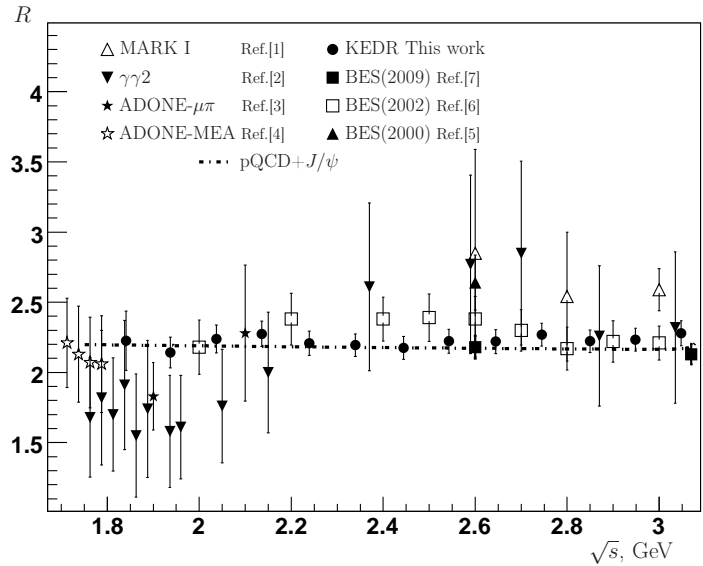}
\end{center}
\caption{{\label{fig:rfinal} The quantity R versus the c.m. energy and 
the sum of the prediction of perturbative QCD and a contribution of the $J/\psi$ resonance.
}}
\end{figure*}

The obtained $R$ values are listed in Table~\ref{tab:rvalues}
and shown in Fig.~\ref{fig:rfinal}.
\begin{table}[ht!]
\caption{\label{tab:rvalues}{Measured values of $R(s)$ with statistical and systematic uncertainties.}}
\begin{center}
\begin{tabular}{lc}      

     $\sqrt{s}$, MeV            & $R(s)$             \\ \hline                
$1841.0 $                & $2.226   \pm 0.139 \pm 0.158$  \\ \hline
$1937.0 $                & $2.141   \pm 0.081 \pm 0.073$  \\ \hline
$2037.3 $                & $2.238   \pm 0.068 \pm 0.072$  \\ \hline
$2135.7 $                & $2.275   \pm 0.072 \pm 0.055$  \\ \hline
$2239.2 $                & $2.208   \pm 0.069 \pm 0.053$  \\ \hline
$2339.5 $                & $2.194   \pm 0.064 \pm 0.048$  \\ \hline
$2444.1 $                & $2.175   \pm 0.067 \pm 0.048$  \\ \hline
$2542.6 $                & $2.222   \pm 0.070 \pm 0.047$  \\ \hline
$2644.8 $                & $2.220   \pm 0.069 \pm 0.049$  \\ \hline
$2744.6 $                & $2.269   \pm 0.065 \pm 0.050$  \\ \hline
$2849.7 $                & $2.223   \pm 0.065 \pm 0.047$   \\ \hline
$2948.9 $                & $2.234   \pm 0.064 \pm 0.051$  \\ \hline
$3048.1 $                & $2.278   \pm 0.075 \pm 0.048$   \\ 
\end{tabular}
\end{center}
\end{table}

\section{Summary}
We have measured the $R$ values  at thirteen  center-of-mass energies 
between 1.84 and 3.05 GeV.
At most of the energy points, the achieved accuracy is about or better 
than  $3.9\%$ at the systematic uncertainty of $2.4\%$.
The obtained $R$ values are compatible with results of the previous experiments
\cite{Mark1:R1977,GG2:R1979,ADONE:R1981,Bai:1999pk,BES:R2002,BES:R2009} but
provide more detailed information on the $R(s)$ quantity in this energy range.

The weighted average $R = 2.225 \pm 0.020 \pm 0.047$ agrees well with
$R_{\rm pQCD} = 2.18 \pm 0.02$ calculated  according to the pQCD 
expansion~\cite{Baikov:pQCD}  for  $\alpha_{s}(m_{\tau})=0.333\pm0.013$ 
derived from the hadronic $\tau$ decays \cite{Brambilla:2014}. 
The averaging was done by taking into account the partial correlation between
systematic uncertainties for different energy points.\\

\section*{Acknowledgments}
We greatly appreciate the efforts of the staff of VEPP-4M to provide
good operation of the complex during long term experiments.
The authors are grateful to V.~P.~Druzhinin and 
E.~P.~Solodov for useful discussions.

This work was supported by Russian Science Foundation under
project N 14-50-00080. Work related to Monte Carlo generators was
partially supported by Russian Foundation for Basic Research under
grant 15-02-05674.

\end{document}